\documentclass[epj]{svjour}

\usepackage{latexsym}
\usepackage{graphics}
\usepackage{amssymb}
\usepackage{xcolor}
\usepackage{amsmath}
\usepackage{multirow}

\newcommand{\ket} [1] {\vert#1\rangle}
\newcommand{\bra} [1] {\langle#1\vert}

\newcommand{\tr}{\textnormal{Tr}}
\newcommand{\id}{\leavevmode\hbox{\small1\normalsize\kern-.33em1}}
\newcommand{\proj}[1]{|#1\rangle\langle#1|}
\newcommand{\beq}{\begin{equation}}
\newcommand{\eeq}{\end{equation}}

\begin{document}

\title{Engineering a C-Phase quantum gate: optical design and experimental realization}
\author{Andrea Chiuri\inst{1}, Chiara Greganti\inst{1} \and Paolo Mataloni\inst{1,2}
% \thanks{\emph{Present address:} Insert the address here if needed}%
}                     

\institute{Dipartimento di Fisica, Sapienza {Universit\`a} di Roma, Piazzale Aldo Moro 5, I-00185 Roma, Italy
\and
Istituto Nazionale di Ottica (INO-CNR), L.go E. Fermi 6, I-50125 Firenze, Italy
}

\date{Received: date / Revised version: date}

\abstract{A two qubit quantum gate, namely the C-Phase, has been realized by exploiting the longitudinal momentum (i.e. the optical path) 
degree of freedom of a single photon. The experimental setup used to engineer this quantum gate represents an advanced version of the 
high stability closed-loop interferometric setup adopted to generate and characterize 2-photon 4-qubit Phased Dicke states. Some experimental results, dealing with 
the characterization of multipartite entanglement of the Phased Dicke states are also discussed in detail.}

\PACS{
 {42.50.Dv}{Quantum state engineering and measurements} \and
 {03.67.Bg}{Entanglement production and manipulation} \and
 {03.67.Lx}{Quantum computation architectures and implementations} \and
 {42.50.Ex}{Optical implementations of quantum information processing and transfer}
%       {PACS-key}{discribing text of that key}   \and
%       {PACS-key}{discribing text of that key}
     } 

\authorrunning{Chiuri A. et al.}

\maketitle

\section{Introduction}
\label{intro}
Quantum entanglement, defined by E. Schr\"oedinger as ``the characteristic trait of quantum mechanics'', 
represents the key resource for modern quantum information (QI).   
An entangled state shared by two or more separated parties is an essential resource for 
fundamental QI protocols, otherwise impossible with classical systems, such as quantum teleportation \cite{benn93prl}, 
quantum computing \cite{raus01prl}, quantum cryptography \cite{eker91prl} and 
quantum dense coding \cite{benn92prl}.
By using entangled states we can also investigate the nonlocal properties of quantum world \cite{eins35pr,bell64phy}.
Quantum optics represents an excellent experimental test bench for various novel concepts 
introduced within the framework of QI theory. Quantum states of photons 
may be easily and accurately manipulated using linear and nonlinear optical devices and 
measured by efficient single-photon detectors.
 
Many QI tasks and fundamental tests of quantum mechanics deal with a large number of qubits. 
For example, the larger the number of qubits, the stronger the violation of Bell inequalities and 
the computational power of a quantum processor. 
Two approaches may be followed to increase the number of qubits.
By the first one the number of entangled particles is increased 
\cite{sack00nat,zhao03prl,kies05prl,lu07nap}. In this way, multi-qubit 
entangled states are created by distributing the qubits between the
particles so that each of them carries one qubit. 
As a second strategy more than one qubit is encoded in each particle, exploiting different 
degrees of freedom (DOFs) of the photon \cite{barb05pra,vall08prl,gao10nap,barr05prl}.
The entanglement of two photons in different DOFs corresponds to produce a hyperentangled (HE) state.
Compared to multiphoton entangled states, HE states offer important advantages as far as purity 
and generation/detection rate are concerned. 
The paper is organized as follows: we describe the generation of 4-qubit Phased Dicke states based on the hyperentanglement of 2 photons. 
We will discuss the experimental results concerning the measurement of a novel 
class of entanglement witness and we will present the first experimental realization of the C-Phase quantum gate based only on the path DOF of a single photon.

\begin{figure}
% 	\centering
\resizebox{0.5\textwidth}{!}{%
  \includegraphics{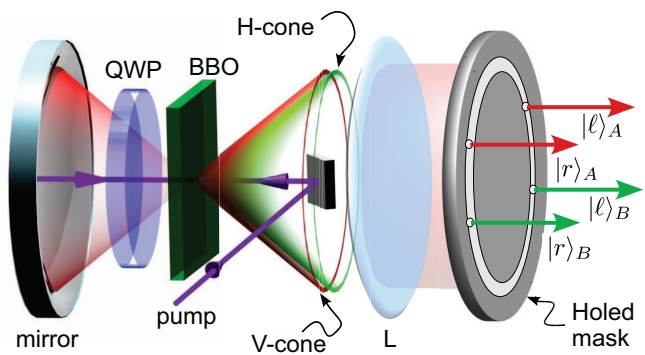}}
\caption{Source of hyperentangled photon states. 
The relative phase between the $\ket{HH}_{AB}$
and $\ket{VV}_{AB}$ contributions can be adjusted by translation of the spherical mirror.
A lens L located at a focal distance from the crystal transforms
the conical emission into a cylindrical one. The dimensionality of the state
can be increased by selecting further pairs of correlated modes on the mask.}
\label{HEsource}
\end{figure}
\section{Hyperentanglement Source}
\label{Sec:HE}
\begin{figure}
% 	\centering
\resizebox{0.5\textwidth}{!}{%
  \includegraphics{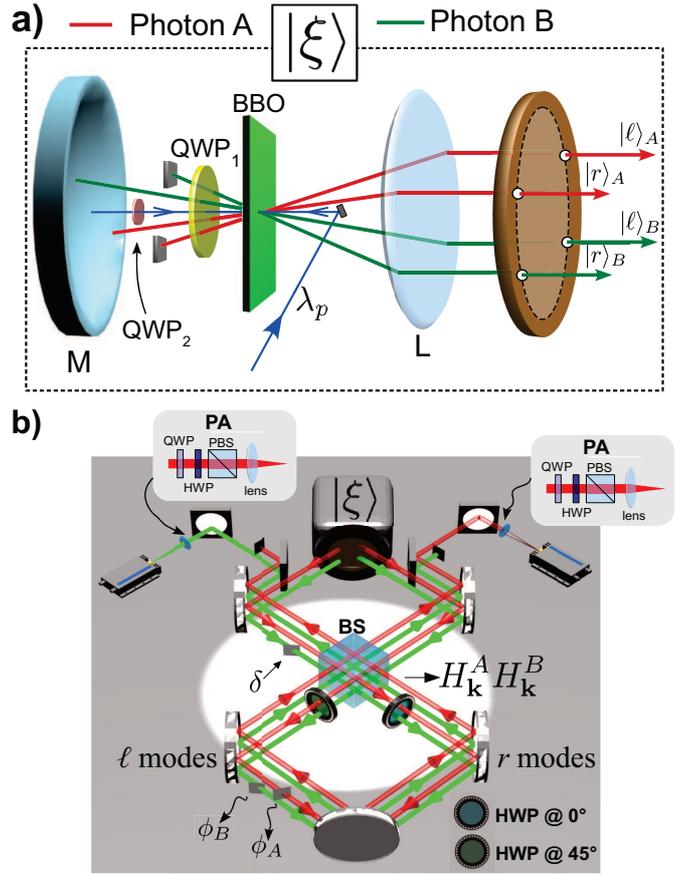}}
\caption{a) Engineered source of the state $\ket{\xi}$. 
The polarization-longitudinal momentum hyperentanglement source has been properly modified to engineer the 
state reported in Eq.(\ref{xi}).
The quarter waveplate $QWP_1$ rotates the polarization of the SPDC photons emitted by the first excitation 
of the crystal while the quarter waveplate $QWP_2$ allows to unbalance the relative weight between the $\ket{HH}$ and the $\ket{VV}$ contributions.
The $\ell$ and $r$ modes on the $V-cone$ are intercepted by two beam stops in order to cancel the 
term $\ket{VV}_{AB}\ket{\ell r}_{AB}$ in the HE state (\ref{HEstate}).
b) Phased Dicke state generation and measurement setup. A thin glass plate, placed before the Sagnac interferometer, 
allows to set the momentum phase $\delta=\pi$. The Phased Dicke state has been obtained by applying 
the Unitary transformation $\mathcal{U}$, shown in Eq.\ref{dickexi}, to the state 
$\ket{\xi}$. The $BS$ allows to implement the Hadamard gates 
in the path DOF while the half waveplate (HWP) at $45^{\circ}$ ($0^{\circ}$), intercepting both the photons, allows to implement the gates
$CX^A_{12}CX^B_{34}$ ($\overline{CZ}^A_{12}\overline{CZ}^B_{34}$). The Pauli operators, in the path DOF, have been
measured by exploiting the second passage through the BS and the thin glass plates $\phi_A$ and $\phi_B$. The necessary measurements
in the polarization DOF have been realized by using an analysis setup, the PA box, before the two detectors. This is composed by HWP, QWP and polarizing beam splitter ($PBS$).}
\label{setupDicke}
\end{figure}
The SPDC source used in this work \cite{cine05lp} is based on the simultaneous entanglement of 2 photons in the 
polarization-longitudinal momentum DOFs. 
The scheme of the source is shown in Fig.\ref{HEsource}. 
Polarization entanglement is created by double excitation (back and forth, after reflection on a
spherical mirror) of a 1 mm Type I BBO crystal by a UV laser beam. The backward emission
determines the so called $V-cone$, with SPDC photon polarization
transformed from horizontal (H) to vertical (V) by double passage of the two photons through a quarter
waveplate (QWP). The forward BBO emission corresponds to the $H-cone$.
Temporal and spatial superposition guarantees indistinguishability of the two emission cones
and allows for the creation of the polarization entangled state
$\frac{1}{\sqrt{2}}(\ket{H}_A \ket{H}_B + e^{i\gamma}\ket{V}_A \ket{V}_B)$, by assuming
the following relations between physical and logical qubits:
$\ket{H}\rightarrow \ket{0}$, $\ket{V}\rightarrow \ket{1}$.

The two photons are emitted with equal probability over symmetrical directions
on the overlapping cone surface then transformed into a cylinder by the lens L [See. Fig.\ref{HEsource}].
By selecting different pairs of correlated emission modes with single mode fibers 
\cite{ross09prl} or with a 4-hole screen \cite{cine05prl} path- (longitudinal momentum-) entanglement is created.
In our experiment, the state $\frac{1}{\sqrt{2}}(\ket{r}_A \ket{\ell}_B + e^{i\delta}\ket{\ell}_A \ket{r}_B)$ 
has been generated by selecting 2 pairs of correlated modes. Here $\ket{r}$ ($\ket{\ell}$)
stands for the optical path followed by the photons in the right (left) direction, with the following relation between physical states and logical qubits, 
$\ket{r}\rightarrow \ket{0}$, $\ket{\ell}\rightarrow \ket{1}$.
The obtained HE state is written as follows:
\begin{equation}\label{HEstate}
\ket{HE_{4}}=\frac{1}{2}(\ket{HH}_{AB} + e^{i\gamma}\ket{VV}_{AB}) \otimes (\ket{r \ell}_{AB}  + e^{i\delta}\ket{\ell r}_{AB} )
\end{equation}
The above described scheme has been also used to explore a higher-dimensional Hilbert space \cite{vall09pra,vall10pra,cecc09prl}. 
In the following we'll describe the use of this setup to generate and measure Phased Dicke states.

\section{Hyperentangled Phased-Dicke states: generation and characterization}
% A Dicke state is a totally symmetric quantum state. In the computational basis \{$\ket{0}$, $\ket{1}$\}, 
% a N-qubit Dicke state with M excitations is given by the coherent superposition of equally weighted permutations of the N-qubit product state as follows:
% \begin{equation}
% \ket{D^{(M)}_{N}}=\frac{1}{\sqrt{C^{(M)}_{N}}}\sum_l{\hat{P}_{l}\ket{0_{0}........0_{N-M}1_{N-M+1}.......1_{N}}}
% \end{equation}
% where $C^{(M)}_{N}$ is the binomial coefficient $\binom{N}{M}$ and $\hat{P}_{l}$'s represent all the possible permutations of the 0's and 1's. 
% Every term of the sum has M logic $\ket{1}$ (called ''excitations'') and N-M logic $\ket{0}$.

In the computational basis \{$\ket{0}$, $\ket{1}$\}, the 4-qubit Phased Dicke state with 2 excitations (i.e. 2 logic $\ket{1}$) is defined as follows:
\begin{eqnarray}
\ket{D_4^{ph}}_{1234}= &&\frac{1}{\sqrt{6}} \big(|0011\rangle + 
 |1100\rangle + |0110\rangle + |1001\rangle \nonumber \\
 &&-|0101\rangle - |1010\rangle \big)_{1234}
% \ket{D^{(ph)}_{4}}=\frac{1}{\sqrt{6}}(\ket{0011}+\ket{1100}+\ket{1001}+\ket{0110}-\ket{1010}-\ket{0101})
\end{eqnarray} 
and derives from the 4-qubit symmetric Dicke state $\ket{D^{(2)}_{4}}_{1234}$ \cite{dicke54pr} by simple unitary transformations: $\ket{D^{(2)}_{4}}_{1234}=Z_{1}Z_{3}\ket{D^{ph}_{4}}_{1234}$.

Dicke states, which have recently attracted much interest for their multipartite entanglement 
properties, have been engineered in multi-photon experiments \cite{kies07prl,prev09prl} while the Phased Dicke states have been
engineered in the hyperentanglement framework \cite{chiu10prl}.

The latter have been obtained by applying suitable unitary transformations on the 2-photon 4-qubit HE states. 
This technique makes possible the 
realization of such multipartite states, 
with relevant advantages in terms of generation rate and state fidelity compared to 4-photon states. 
The measurements were performed by a {\it closed-loop Sagnac} scheme with intrinsic almost perfect stability.

\subsection{State generation} 
Here we briefly describe how the experimental setup of Fig.\ref{setupDicke} has been used in ref.\cite{chiu10prl} to engineer Phased Dicke states. 
Let us consider the following state 
$\ket{\xi}_{1234}\equiv\frac{1}{\sqrt{6}}(\ket{0010}-\ket{1000}+2\ket{0111})_{1234}$.
The Phased Dicke state can be obtained by
applying a unitary transformation 
 $\mathcal U$ to the state $\ket \xi$:
\begin{equation}
{\ket{D^{(ph)}_{4}}_{1234}=Z_4\overline{CZ}_{12}\overline{CZ}_{34}CX_{12}CX_{34}H_1H_3\ket{\xi}\equiv\mathcal U\ket{\xi}}_{1234}
\label{dickexi}
\end{equation}
where $H_j$ and $Z_j$ stands for the Hadamard and the Pauli $\sigma_z$ 
transformations on qubit $j$, $CX_{ij}=\ket{0}_i\bra{0}\id_j+\ket1_i\bra1X_j$ 
is the controlled-NOT gate and $\overline{CZ}_{ij}=\ket1_i\bra1\id_j+\ket0_i\bra0Z_j$ 
the controlled-Z [see Fig.\ref{setupDicke}]. 
The transformations $\overline{CZ}_{12}\overline{CZ}_{34}$ are needed to compensate the optical delay introduced by the $CX$ gates
in the Sagnac loop of Fig. \ref{setupDicke}b).
As explained in the previous Section, the $\ket{0}$ and $\ket{1}$ states are encoded into
horizontal $\ket{H}$ and vertical $\ket{V}$ polarization or into
right $\ket{r}$ and left $\ket{\ell}$ path. The qubit 1 (2) belongs to the path (polarization) DOF of the photon A while 
the qubit 3 (4) belongs to the path (polarization) DOF of the photon B.

According to those relations the state $\ket{\xi}$  reads:
\begin{equation}\label{xi}
\begin{aligned}
\ket{\xi}_{1234}=\frac{1}{\sqrt{6}}[(\ket{r\ell}-\ket{\ell r})_{13}\ket{HH}_{24}+
2\ket{r\ell}_{13}\ket{VV}_{24}]
\end{aligned}
\end{equation}
and may be obtained by suitably modifying the source used to realize polarization-longitudinal momentum
hyperentangled states \cite{barb05pra,cecc09prl} (see Sec.\ref{Sec:HE}).
Let us consider now the HE state in Eq.(\ref{HEstate}) and the Fig.\ref{setupDicke}a). 
The SPDC contribution, due to the pump beam incoming after reflection on mirror $M$, corresponds to the term 
$\ket{HH}(\ket{r\ell}-\ket{\ell r})$, whose weight is determined by the
quarter waveplate $QWP_{2}$ intercepting the UV beam (see \cite{vall07pra} for more details on the generation of 
the non-maximally polarization entangled state).
The other SPDC contribution $2\ket{VV}\ket{r\ell}$ is determined
by the first excitation of the pump beam: 
here the $\ket{\ell r}$ modes are intercepted by two beam stops and the 
quarter waveplate $QWP_{1}$ transforms the $\ket{HH}$ SPDC emission into $\ket{VV}$ after reflection on mirror $M$. 
The relative phase between the $\ket{VV}$ and $\ket{HH}$ is varied by
translation of the spherical mirror M. 

The transformation \eqref{dickexi} $\ket{\xi}\rightarrow\ket{D^{(ph)}_{4}}$ 
is realized by using two waveplates and one beam splitter (BS): 
the two Hadamards $H_1$ and $H_3$ in \eqref{dickexi},
acting on both path qubits, are implemented by a single BS
for both A and B modes. For each controlled-NOT (or controlled-Z) gate appearing 
in \eqref{dickexi} the control and target qubits
are respectively represented by path and polarization of a single photon:
a half waveplave (HWP) with axis oriented at 45$^\circ$ (0$^\circ$) with respect to the vertical direction
and located into the left $\ket{\ell}$ (right $\ket r$) mode implements a $CX$ ($\overline{CZ}$) gate.

After these transformations, 
the optical modes are spatially matched the second time on the BS, 
closing in this way a {\it closed-loop Sagnac} interferometer that allows
high stability in measuring the path Pauli operators [see Fig. \ref{setupDicke}b)].
Polarization Pauli operators are measured
by standard polarization analysis setup 
in front of detectors (i.e. PA box in Fig.\ref{setupDicke}b)).

Note that, the $\ket{0}$ ($\ket{1}$) state, for the path DOF,
is identified by the clockwise (counterclockwise) mode in the Sagnac loop.

It is worth of stressing once more the high stability guaranteed by the Sagnac 
interferometric scheme in performing the path analysis . 

\subsection{Entanglement characterization via structural witness} 
The presence of entanglement in the generated Phased Dicke states was tested by adopting a recently proposed 
class of entanglement witnesses, so-called structural witnesses \cite{kram09prl}. 

For a composite system of $N$ particles, the structural witnesses 
\cite{kram09prl} have the form
\begin{equation} 
\label{genentwit}
W(\vec{k}) := \id_N - \Sigma(\vec{k})
\end{equation}
where $\vec{k}$ is a real parameter (the three dimensional wave-vector transferred in a scattering
scenario), $\id_N$ is the identity operator and
\begin{equation} \label{sigmagennew}
	\Sigma(k^x,k^y,k^z) 
	= \frac{1}{B(N,2)} 
[c_x \hat{S}^{xx}(k^x) 
+ c_y  \hat{S}^{yy}(k^y) + c_z \hat{S}^{zz}(k^z)],
\end{equation}
% \begin{equation} 
% \label{sigmagen}
% \bar\Sigma(k) = \frac{1}{B(N,2)} 
% [ c_x \hat{S}^{xx}(k) 
% + c_y  \hat{S}^{yy}(k) + c_z \hat{S}^{zz}(k)], 
% \end{equation}
with $c_i \in \mathcal{R}$, $|c_i| \leq 1$.
Here $B(N,2)$ is the binomial coefficient and the structure factor operators
$\hat{S}^{\alpha\beta}(k)$ are defined as 
\begin{equation} \label{Sab}
\hat{S}^{\alpha \beta}(k) := \sum_{i<j} e^{ik(r_i -r_j)}S_i^{\alpha} S_j^\beta,
\end{equation}
where $i,j$ denote the $i$-th and $j$-th spins, $r_i, r_j$ their positions
in a one-dimensional scenario, 
and $S_i^{\alpha}$ are the spin operators with $\alpha,\beta = x,y,z$. 
A suitable structural witness {$\overline{\mathcal W}$} for the Phased Dicke state 
can be constructed by considering $k^x=k^y=\pi$ and $k^z=0$:
\begin{equation} 
\label{wnew}
\overline{\mathcal W} = \id_N -\frac{1}{6}[\hat{S}^{xx}(\pi)+\hat{S}^{yy}(\pi)-
\hat{S}^{zz}(0)]\ .
\end{equation}
The expectation value of the above witness for the Phased Dicke state is given by 
$\tr (\proj{D_4^{ph}} \overline{\mathcal W})=-\frac23$,
thus leading to a robust entanglement detection in the presence of noise.
The witness $\overline{\mathcal W}$ measured for the Phased Dicke state \cite{chiu10prl}, is 
\beq\label{Wexp}
\langle{\overline{\mathcal W}}\rangle_{exp}=-0.382\pm0.012
\eeq
We report in Table \ref{TabMeasur} the experimental 
values for each operator appearing in the Witness (\ref{wnew}).

\begin{table}[ht]       
\centering
\caption{Experimental values of the operators needed to estimate 
the structural witness in Eq.\ref{wnew}. The uncertainties are determined by associating
Poissonian fluctuations to the coincidence counts. Here $k$ refers to the longitudinal momentum DOF 
while $\pi$ refers to the polarization DOF.}
\label{TabMeasur}
\begin{tabular}{cccc}
\hline\noalign{\smallskip}
Operators & Involved & Local & Results\\
 & Qubits & Settings & \\
\noalign{\smallskip}\hline\noalign{\smallskip}
 & $1^{\circ}2^{\circ}3^{\circ}4^{\circ}$ & $(1^{\circ}3^{\circ})k(2^{\circ}4^{\circ})\pi$ & \\
\noalign{\smallskip}\hline\noalign{\smallskip}
$S^{xx}_{14}$ & X11X & (X1)k(1X)$\pi$  & $-0.458\pm0.013$\\
$S^{xx}_{24}$ & 1X1X & (11)k(XX)$\pi$  & $0.531\pm0.012$\\
$S^{xx}_{34}$ & 11XX & (1X)k(1X)$\pi$  & $-0.384\pm0.013$\\
$S^{xx}_{12}$ & XX11 & (X1)k(X1)$\pi$  & $-0.545\pm0.012$\\
$S^{xx}_{13}$ & X1X1 & (XX)k(11)$\pi$  & $0.597\pm0.011$\\
$S^{xx}_{23}$ & 1XX1 & (1X)k(X1)$\pi$  & $-0.620\pm0.011$\\
$S^{yy}_{14}$ & Y11Y & (Y1)k(1Y)$\pi$  & $-0.617\pm0.009$\\
$S^{yy}_{24}$ & 1Y1Y & (11)k(YY)$\pi$  & $0.590\pm0.009$\\
$S^{yy}_{34}$ & 11YY & (1Y)k(1Y)$\pi$  & $-0.528\pm0.009$\\
$S^{yy}_{12}$ & YY11 & (Y1)k(Y1)$\pi$  & $-0.550\pm0.009$\\
$S^{yy}_{13}$ & Y1Y1 & (YY)k(11)$\pi$  & $0.523\pm0.010$\\
$S^{yy}_{23}$ & 1YY1 & (1Y)k(Y1)$\pi$  & $-0.425\pm0.010$\\
$S^{zz}_{14}$ & Z11Z & (Z1)k(1Z)$\pi$  & $-0.327\pm0.024$\\
$S^{zz}_{24}$ & 1Z1Z & (11)k(ZZ)$\pi$  & $-0.304\pm0.024$\\
$S^{zz}_{34}$ & 11ZZ & (1Z)k(1Z)$\pi$  & $-0.314\pm0.024$\\
$S^{zz}_{12}$ & ZZ11 & (Z1)k(Z1)$\pi$  & $-0.354\pm0.024$\\
$S^{zz}_{13}$ & Z1Z1 & (ZZ)k(11)$\pi$  & $-0.308\pm0.024$\\
$S^{zz}_{23}$ & 1ZZ1 & (1Z)k(Z1)$\pi$  & $-0.315\pm0.024$\\
\noalign{\smallskip}\hline
\end{tabular}
% Or use
% \vspace*{15cm}  % with the correct table height
\end{table}

\begin{figure*}[ht]
\centering
\resizebox{0.85\textwidth}{!}{%
  \includegraphics{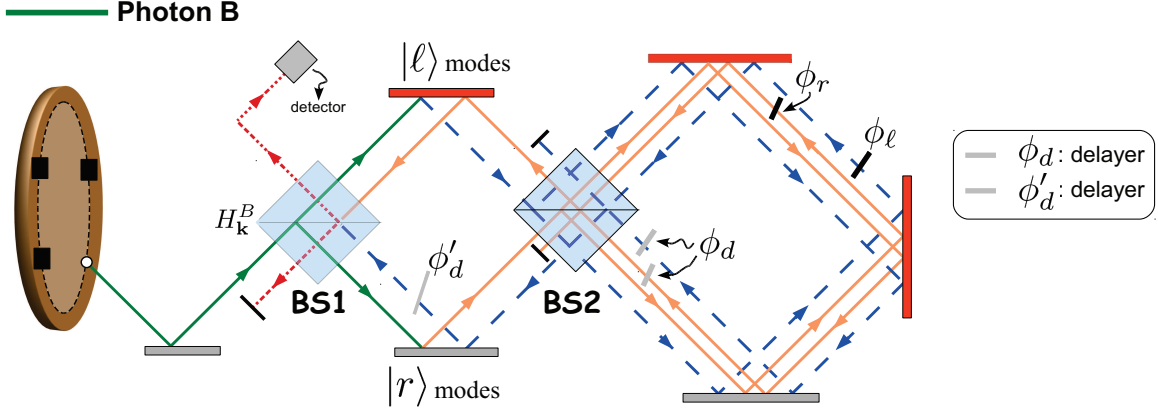}}
\caption{C-Phase gate experimental setup based only on the path DOF of a single photon. 
The control qubit is identified by the different paths followed by the photon after the $BS1$ (i.e. $\ket{r}$ or $\ket{\ell}$),
while the target qubit is given by the clockwise ($\ket{C}$) or counterclockwise ($\ket{A}$) path followed by the photon after $BS2$ within the displaced Sagnac
 interferometer.
The phase shift performed by the gate has been obtained by using the two thin glass plates $\phi_{\ell}$ and $\phi_{r}$, both 
on the counterclockwise paths $\ket{A}$.
Two delayers $\phi_{d}$ allow to compensate the temporal delay
introduced by $\phi_{\ell}$ and $\phi_{r}$. The insertion of $\phi'_{d}$ is needed to avoid 
interference between the modes coming back from the displaced Sagnac system and impinging on $BS1$.}
\label{SetupCphase}
\end{figure*}

We have also measured a witness $W_{mult}$, introduced in \cite{toth09njp},
to demonstrate the genuine multipartite nature of the generated state.
This operator is defined as follows:
\begin{equation}
W_{mult}= 2\cdot\id + \frac{1}{6}(\hat{J}^{2}_{x}+\hat{J}^{2}_{y}-\hat{J}^{4}_{x}-\hat{J}^{4}_{y})+ \frac{31}{12}\hat{J}^{2}_{z}-\frac{7}{12}\hat{J}^{4}_{z}
\end{equation}
where $\hat{J}^{2}_{i}=1+\frac{1}{2}\hat{S}^{ii}(k^{i})$ and $\hat{J}^{4}_{i}=1+\hat{S}^{ii}(k^{i})+\frac{1}{4}(\hat{S}^{ii})^{2}(k^{i})$, i=x,y,z and $k^{x}=k^{y}=\pi$, $k^{z}=0$.
It comes out that this equation, in terms of the operators $\hat{S}^{ii}(k^{i})$ defined in Eq.\ref{Sab}, reads:
\begin{eqnarray} 
W_{mult}=&&\frac{1}{8}(2\cdot\id-2\hat{S}^{xx}(\pi)-2\hat{S}^{yy}(\pi)+\hat{S}^{zz}(0)\nonumber\\
-&&7\hat{S}^{zzzz}-2\hat{S}^{xxxx}-2\hat{S}^{yyyy})
\end{eqnarray}
with $\hat{S}^{zzzz}=Z_{1}Z_{2}Z_{3}Z_{4}$, $\hat{S}^{xxxx}=X_{1}X_{2}X_{3}X_{4}$, $\hat{S}^{yyyy}=Y_{1}Y_{2}Y_{3}Y_{4}$, here the subscripts indicate the qubits 
involved in the measurement.
The measured values of the operators $\hat{S}^{ii}(k^{i})$ are reported in Table \ref{TabMeasur}.
By taking into account also the results reported in Table \ref{4operator}, we obtained 
\begin{equation}
\langle W_{mult}\rangle=-0.341\pm0.015
\end{equation}
These results already appeared in \cite{chiu10prl}, where a detailed discussion of the 
experimental results was lacking. 
In the next Section we will describe how the same experimental setup of Fig.\ref{setupDicke}, 
properly modified, has been adopted to realize a single photon C-phase gate. 

\begin{table}
\centering
\caption{Experimentally measured expectation values of collective
spin operators for the Phased Dicke state. 
The uncertainties are determined by associating
Poissonian fluctuations to the coincidence counts.}
\label{4operator}       
\begin{tabular}{ccc}
\hline\noalign{\smallskip}
Operators & Local Settings & Results\\
\noalign{\smallskip}\hline\noalign{\smallskip}
$1^{\circ}2^{\circ}3^{\circ}4^{\circ}$ & $(1^{\circ}3^{\circ})k(2^{\circ}4^{\circ})\pi$ & \\
\noalign{\smallskip}\hline\noalign{\smallskip}
$X_{1}X_{2}X_{3}X_{4}$ & (XX)k(XX)$\pi$  & $0.673\pm0.011$\\
$Y_{1}Y_{2}Y_{3}Y_{4}$ & (YY)k(YY)$\pi$  & $0.635\pm0.009$\\
$Z_{1}Z_{2}Z_{3}Z_{4}$ & (ZZ)k(ZZ)$\pi$  & $0.922\pm0.010$\\
\noalign{\smallskip}\hline
\end{tabular}
% Or use
%\hspace*{5cm}  % with the correct table height
\end{table}

\section{Experimental realization of the C-Phase quantum gate}
\label{Seccphase}
Many efforts have been made in the last years to experimentally implement several 
basic quantum gates, such as the CNOT or C-Phase gate. The latter was 
in particular realized by exploiting the polarization DOF of a photonic system \cite{2kies05prl}
and, more recently, was implemented within a quantum dot scenario \cite{meun11prb}.
The unitary transformation corresponding to the C-Phase, is defined as follows:
\begin{equation}
{\mathcal U^{phase}}=
\begin{pmatrix}
1 & 0  &  0  &  0\\

0  & e^{\phi_{1}}  &  0  &  0\\

0  &  0  &  1 &  0\\

0  &  0  &  0  &  e^{\phi_{2}}

\end{pmatrix}
\label{transf}
\end{equation}
\begin{figure*}[ht]
\centering
\resizebox{0.85\textwidth}{!}{%
  \includegraphics{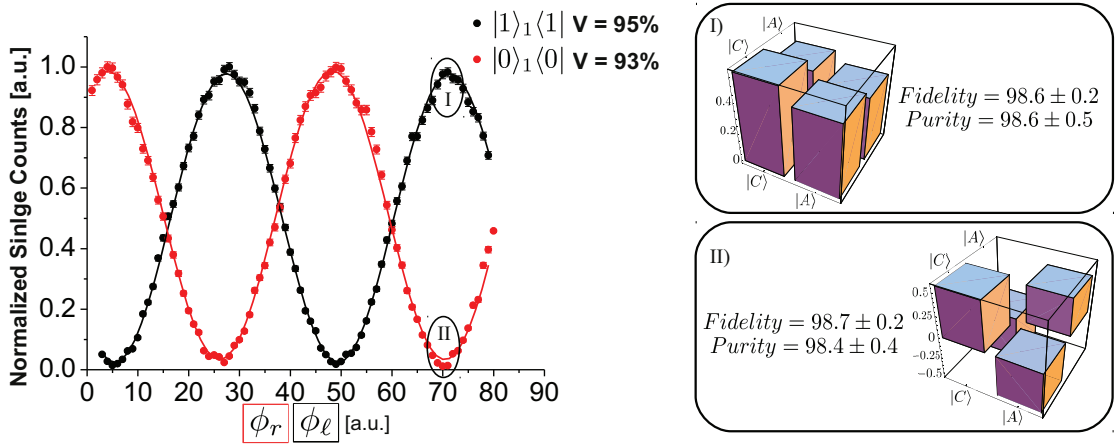}}
\caption{Measured oscillations of the single counts with dots representing the experimental data and the solid line corresponding to the fitting curve. 
The dark counts have been subtracted.
The uncertainties have been determined by associating Poissonian fluctuations to the single counts.
The red (black) data have been measured by projecting the state reported in Eq. (\ref{input}) on $\ket{0}_1 \bra{0}$ ($\ket{1}_1 \bra{1}$)
and varying $\phi_{r}$ ($\phi_{\ell}$) with the thin glass plates in the displaced Sagnac interferometer. For $\phi_{r}=\pi$ and $\phi_{\ell}=0$ we performed the Quantum
State Tomography of qubit 2. The Fidelity have been calculated with respect to the theoretical ones, i.e. $\ket{+}_2 \bra{+}$ for the state in the box I and $\ket{-}_2 \bra{-}$ 
for the state in the box II.}
\label{results}
\end{figure*}
The optical setup of Fig.\ref{SetupCphase} shows the high stability {\it closed-loop displaced Sagnac} scheme 
used in the experiment.
It represents a modified version of the one adopted for the Phased Dicke state experiment. 
Here a second beam splitter ($BS2$) intercepting only the optical path of lower photon has been added. 
The particular position of the $BS2$ enables the realization of a {\it diplaced Sagnac} interferometer, i.e. an 
interferometric scheme where the right mode $\ket{r}$ 
and the left mode $\ket{\ell}$ impinge the $BS2$ in different points.

Let us now describe how the implemented gate works.
In the HE source, described in Sec.\ref{Sec:HE}, only one polarization cone, namely the $H-cone$, is considered and
only one mode, corresponding to the lower photon, is taken into account.
In order to explain the experiment let us consider only the $\ket{r}_B$ mode 
coming out of the holed mask, as reported in Fig.\ref{SetupCphase}.
The $BS1$ acts as follows:
\begin{equation}\label{BS1}
\ket{r}_B \xrightarrow{BS1} \frac{1}{\sqrt{2}}(\ket{r}_B + \ket{\ell}_B).
\end{equation}
The photon, arriving at the $BS2$, can go clockwise ($\ket{C}_B$) or counterclockwise ($\ket{A}_B$) within the {\it diplaced Sagnac}. This
corresponds to add a further qubit, encoded in the path DOF, hence the state in Eq.(\ref{BS1}) becomes:
\begin{equation}\label{input2}
\frac{1}{\sqrt{2}}(\ket{r}_B + \ket{\ell}_B) \xrightarrow{BS2} \frac{1}{\sqrt{2}}(\ket{r}_B\ket{\phi_{r}}_B + \ket{\ell}_B\ket{\phi_{\ell}}_B)
\end{equation}
where $\ket{\phi_{r}}_B=\frac{1}{\sqrt{2}}(\ket{C}_B+ e^{i\phi_{r}}\ket{A}_B)$,  $\ket{\phi_{\ell}}_B=\frac{1}{\sqrt{2}}(\ket{C}_B+ e^{i\phi_{\ell}}\ket{A}_B)$.
By considering the following relations between logical states and physical qubits:
\begin{eqnarray}
&&\{\ket{0}_{1},\ket{1}_{1}\}\rightarrow\{\ket{r}_{B},\ket{\ell}_{B}\}\nonumber \\
&&\{\ket{0}_{2},\ket{1}_{2}\}\rightarrow\{\ket{C}_{B},\ket{A}_{B}\}
\end{eqnarray}
the state (\ref{input2}) reads:
\begin{eqnarray}\label{input}
\frac{1}{2}&&[\ket{0}_1 \otimes(\ket{0}+e^{\phi_{r}}\ket{1})_2 + \ket{1}_1 \otimes (\ket{0}+e^{\phi_{\ell}}\ket{1})_2 ]=\nonumber \\
\frac{1}{2}&&(\ket{00}_{12} + e^{\phi_{r}}\ket{01}_{12}  +\ket{10}_{12} + e^{\phi_{\ell}}\ket{11}_{12} )
\end{eqnarray}
The phases $\phi_{r}$ and $\phi_{\ell}$ can be indipendently varied by using 
two thin glass plates placed within the interferometer. 
This corresponds to implement the transformation reported in Eq.(\ref{transf}) with $\phi_{r}=\phi_{1}$ and $\phi_{\ell}=\phi_{2}$.
It is worth to remember that both the control and target qubits of the quantum gate are encoded in the path DOF of 
photon B. Precisely, the control qubit is encoded in the longitudinal momentum of the photon before 
$BS2$ (i.e. \{$\ket{r}_B$,$\ket{\ell}_B$\}) while the target qubit is encoded in the path followed in the Sagnac 
scheme (i.e. \{$\ket{C}_B$,$\ket{A}_B$\}).
We report in Table \ref{truth} the ``truth table'' of the engineered gate.
\begin{table}[ht]     
\centering
\caption{``Truth table'' of the realized C-phase gate. In the first column we report
the logical qubits while in the second column there are the corresponding 
physical qubits.}
\label{truth}  
\begin{tabular}{|cc|cc|}
\hline
\multicolumn{2}{|c|}{Logical qubit} & \multicolumn{2}{|c|}{Physical qubit}\\ \hline
Control & Target & Control & Target\\ \hline
$\ket{0}_1\bra{0}$ & $\frac{1}{\sqrt{2}}(\ket{0}_2+ e^{i\phi_{r}}\ket{1}_2)$ & $\ket{r}_B\bra{r}$ & $\frac{1}{\sqrt{2}}(\ket{C}_B+e^{\phi_{r}}\ket{A}_B)$ \\
$\ket{1}_1\bra{1}$ & $\frac{1}{\sqrt{2}}(\ket{0}_2+ e^{i\phi_{\ell}}\ket{1}_2)$ & $\ket{\ell}_B\bra{\ell}$ & $\frac{1}{\sqrt{2}}(\ket{C}_B+e^{\phi_{\ell}}\ket{A}_B)$\\ \hline
\end{tabular}
\end{table}

The second passage through $BS2$ allows to perform the measurement of the Pauli operators.

The obtained experimental results are shown in Fig.\ref{results}.
We measured the oscillations of the single counts by projecting the state (\ref{input}) on $\ket{0}_1 \bra{0}$ ($\ket{1}_1 \bra{1}$) and varying        
$\phi_{r}$ ($\phi_{\ell}$). The projection on $\ket{r}_B \bra{r}$ ($\ket{\ell}_B \bra{\ell}$) was performed by intercepting the input mode 
$\ket{\ell}_B$ ($\ket{r}_B$).\\ In the experiment, $\phi_{r}$ = $\phi_{\ell}+\pi$, thus there is a particular phase factor between $\phi_{r}$ and $\phi_{\ell}$, 
however it is important to underline that they can assume any general value with this setup. In the case $\phi_{\ell}=0$, $\phi_{r}=\pi$, we have performed the 
tomographic reconstruction \cite{jame01pra} of the density matrix related to the state $\ket{\phi_{r}}_B\bra{\phi_{r}}$ and $\ket{\phi_{\ell}}_B\bra{\phi_{\ell}}$. 
These values correspond to realize a $C-NOT$ gate.
As already pointed out, the second passage through $BS2$ allows to measure the Pauli operators $\hat{\sigma_x}$ and $\hat{\sigma_y}$. 
The third Pauli operator $\hat{\sigma_z}$ has been measured by intercepting the mode in the displaced Sagnac (i.e. $\ket{C}$ or $\ket{A}$). 
This corresponds to make a projection on the computational basis. The fidelities of the measured states, calculated with respect to the theoretical states, are larger 
than 98\%. 
\begin{figure}[ht]
\centering
\resizebox{0.50\textwidth}{!}{%
  \includegraphics{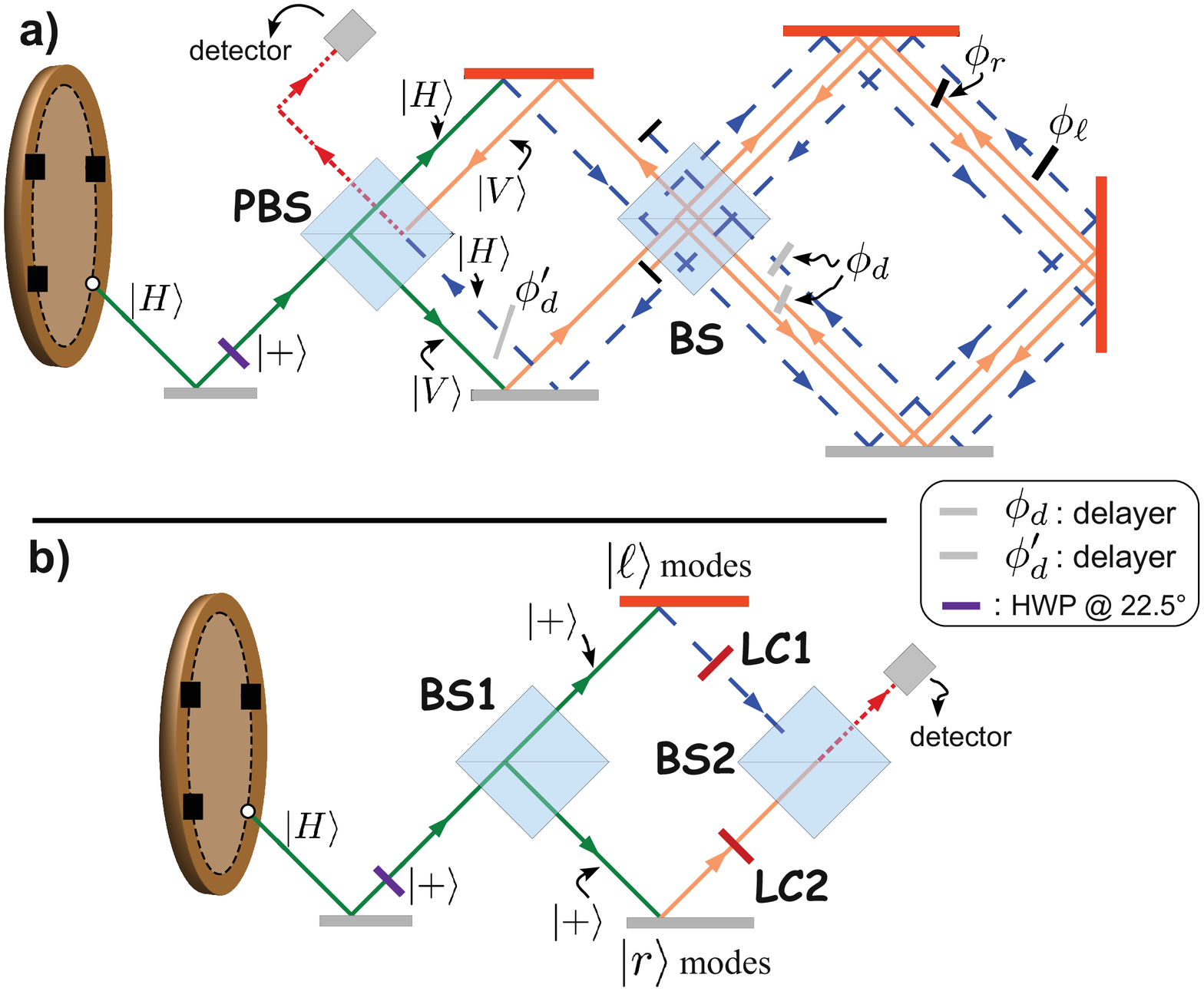}}
\caption{Two alternative schemes allowing the realization of the C-Phase gate based on two 
DOFs of a single photon, the polarization and the path.
a) The photon entering the $PBS$ is in the state $\ket{+}=\frac{1}{\sqrt{2}}(\ket{H}+\ket{V})$ and the $PBS$ separates the 
two polarizations $\ket{H}$ and $\ket{V}$. The displaced Sagnac acts as already explained in the Sec.\ref{Seccphase} and the two modes 
coming back to the $PBS$ are sent towards the same detector. b)  The photon entering the BS1 is in the polarization state 
$\ket{+}=\frac{1}{\sqrt{2}}(\ket{H}+\ket{V})$. After the BS1 the photon is in the state $\frac{1}{\sqrt{2}}(\ket{r}_{k}\ket{+}_{\pi}+\ket{\ell}_{k}\ket{+}_{\pi})$, here 
the subscript $\pi$ ($k$) indicates the polarization (path) DOF. The employment of two liquid crystals (LC1 and LC2) allows to 
vary the relative phase between the polarizations $\ket{H}$ and $\ket{V}$. Precisely, these values can be independently set for the 
$\ket{\ell}$ and $\ket{r}$ modes.}
\label{Cphase2}
\end{figure}
\section{Conclusions and Discussion}
In this work we have presented the main features of a 4-qubit Phased Dicke state, 
built on the polarization and longitudinal momentum of 
the photons. The entanglement properties have been investigated by a new kind of entanglement 
witness, so-called structural witness. To generate and measure this state, an interferometric 
{\it closed-loop Sagnac} scheme with almost perfect intrinsic stability has been adopted.
An advanced version of this setup has allowed to efficiently 
implement the C-Phase quantum gate based on the 
optical path of a single photon. We have presented the obtained experimental results and discussed the flexibility showed by the engineered setup.   

Other experimental schemes can be conceived to realize such quantum gate. 
For instance two changes can be implemented [See Fig.\ref{Cphase2} a)]: 
\begin{itemize}
\item by replacing the $BS1$ with a $PBS$ 
\item by exploiting the polarization of the photon before it arrives at the $BS1$. Precisely it has to be in the state 
$\ket{+}=\frac{1}{\sqrt{2}}(\ket{H}+\ket{V})$ and this can be obtained by placing a half-waveplate rotated by $22.5^{\circ}$ with respect to the vertical polarization.
\end{itemize}
In this case, the $PBS$ will separate the polarization $H$ and $V$ and the displaced Sagnac will act as already explained in the previous section. 
In this case, the modes coming back to the $PBS$ will be sent towards the same detector \footnote{The horizontal polarization is tansmitted while 
the vertical polarization is reflected.}.\\
Another possibility, sketched in Fig.\ref{Cphase2}b), concerns the use of path DOF as the control qubit and of
polarization DOF as the target. Let us consider the input photon in the state $\ket{+}$ encoded in the polarization DOF. Depending on the optical 
path followed after the $BS1$, an arbitrary phase can be experimentally assigned to the polarization state by employing liquid crystals \cite{chiu11prl}. 

Recent developments of integrated quantum circuits suggest to adopt these systems to realize an intrinsically stable C-Phase gate based on path encoded qubits. 
It has been recently demonstrated that, due to the low birefringence, integrated quantum circuits written by femtosecond laser pulses can support polarization 
qubits \cite{sans10prl,sans12prl,sans11nat}. Hence, using this approach to implement the C-phase gate demonstrated in this experiment and the proposed schemes 
sketched in Fig.\ref{Cphase2}a) may open interesting developments in a very challenging research field.

% 
% The engineered setup has also another considerable feature. We have already pointed out that the use of the displaced Sagnac has been necessary in order 
% to guarantee a high stability. The integrated quantum circuit recently realized \cite{sans10prl,sans12prl} have an intrinsic stability and are 
% polarization insensitive. For these reasons, the realized experiment and the 
% proposed scheme, sketched in Fig.\ref{Cphase2}a), can be implemented also with the integrated quantum circuits, thus they 
% could represent interesting developments in a very challenging research field.

%\begin{figure}
%% Use the relevant command for your figure-insertion program
%% to insert the figure file.
%% For example, with the option graphics use
%\resizebox{0.75\textwidth}{!}{%
%  \includegraphics{leer.eps}
%}
%% If not, use
%%\vspace{5cm}       
%\label{fig:1}    
%\end{figure}

%
% BibTeX users please use
% \bibliographystyle{}
% \bibliography{}

\end{document}